\documentclass{svjour3}                     
\usepackage{graphicx}

\usepackage[utf8]{inputenc}
\usepackage{amsmath}
\usepackage{amsfonts}
\usepackage{amssymb}
\usepackage{breakcites}
\usepackage[strings]{underscore}
\usepackage{natbib}

\begin{document}

\title{Convergence tests for transdimensional Markov Chains in geoscience imaging
}


\author{Márk Somogyvári         \and
        Sebastian Reich 
}


\institute{M. Somogyvári \and S.Reich\at 
              Institute of Mathematics, University of Potsdam, D-14476 Potsdam, Germany\\
              \email{mark.somogyvari@uni-potsdam.de}           
}

\date{Published in Mathematical Geosciences: 20 May 2019}

\maketitle

\begin{abstract}
Classic inversion methods adjust a model with a predefined number of parameters to the observed data. With transdimensional inversion algorithms such as the reversible-jump Markov Chain Monte Carlo (rjMCMC), it is possible to vary this number during the inversion and to interpret the observations in a more flexible way. Geoscience imaging applications use this behaviour to automatically adjust model resolution to the inhomogeneities of the investigated system, while keeping the model parameters on an optimal level.
The rjMCMC algorithm produces an ensemble as result, a set of model realizations which together represent the posterior probability distribution of the investigated problem. The realizations are evolved via sequential updates from a randomly chosen initial solution, and converge toward the target posterior distribution of the inverse problem. Up to a point in the chain, the realizations may be strongly biased by the initial model, and have to be discarded from the final ensemble. With convergence assessment techniques, this point in the chain can be identified.
Transdimensional MCMC methods produce ensembles which are not suitable for classic convergence assessment techniques because of the changes in parameter numbers. To overcome this hurdle, three solutions are introduced to convert model realizations to a common dimensionality while maintaining the statistical characteristics of the ensemble. A scalar, a vector and a matrix representation for models is presented, inferred from tomographic subsurface investigations, and three classic convergence assessment techniques are applied on them. It is shown that appropriately chosen scalar conversions of the models could retain similar statistical ensemble properties as geologic projections created by rasterization.
\end{abstract}

\section{Introduction}
\label{intro}
Transdimensional inversion methods, such as the reversible-jump Markov Chain Monte Carlo algorithm (rjMCMC), are novel tools of Bayesian statistics that can be used to solve inverse problems where the number of parameters (the dimensionality of the model) is unknown before the inversion. These algorithms adjust the model dimensions automatically, giving extra flexibility to the inversion compared to classical approaches. The most common transdimensional inversion method is the rjMCMC by \citet{Green1995}, which has been tested out in many applications of geosciences in the recent years \citep{Bodin2012,Dettmer2012,Saygin2016,Fox2015,Mondal2010,Jimenez2016}.\\
MCMC methods create Markov chains with a stationary distribution approximating the posterior probability distribution of the investigated problem. This process is also called MCMC sampling, as the algorithm takes samples from the posterior to produce a set of samples that represents the complete distribution. The samples are taken via an iterative random search, where subsequent elements of the Markov chain are generated via random perturbations \citep{Robert1998}. \\
The rjMCMC algorithm or Metropolis-Hastings-Green algorithm is the extended (generalized) version of classic MCMC methods \citep{Brooks2011}, with updates that introduce new parameters or remove existing ones from the models. The transdimensional updates are then evaluated by the Metropolis-Hastings-Green acceptance criteria, ensuring that the stationarity of the Markov chain is preserved.
The algorithm results in an ensemble, a large set of model realizations that provide a mapping of the posterior distribution. Thus, these methodologies yield more than one calibrated model, and they also provide information on the uncertainty of the estimate.
In the case of transdimensional inversion, the posterior probability function spans over different dimensionalities, as models within the ensemble have different numbers of parameters.\\
This makes the calculation of the statistical properties of transdimensional ensembles problematic, as it is not possible to follow each parameter along the complete Markov chain. The most important aspect of ensemble analysis would be convergence assessment, quantifying the quality of the rjMCMC chain (convergence, independence of samples) and if it requires more iterations. MCMC simulations typically start from a randomly chosen initial model and converge to the stationary distribution of the posterior of the inverse problem. Samples from the convergence phase can be biased by the choice of the initial model and they have to be discarded from the ensemble (they are called the burn-in of the Markov chain). Finding the point of convergence within a chain is especially important in MCMC applications where iterations are computationally expensive and only limited chain lengths are achievable (conceptually infinite long chains would provide the perfect sampling of the posterior). In these situations, knowing the convergence point more accurately could mean that fewer models have to be discarded, and a longer part of the chain could be kept as the ensemble.\\
Classic convergence assessment tools try to find the point of convergence by testing where the chain has become stationary \citep{Cowles1996a}. Their application to transdimensional chains is not possible directly, because they rely on the calculation of the statistical properties of each individual parameter and they are not designed to handle changes in their numbers \citep{Bodin2009}. This is one of the factors that makes the rjMCMC method less popular for geoscience applications than classic MCMC techniques, that are  extensively used \citep{Jeong2017,Elsheikh2012,OMalley2018,DjibrillaSaley2016,AfshariMoein2018}\\
These restrictions were encountered in the authors previous work \citep{Somogyvari2017}, where no quantitative convergence assessment was done. This gives the main motivation for this study and the inversion algorithm presented in that paper is used here. The main goal is to provide universal diagnostic methods that work without any major modification of the existing inversion algorithms. For this purpose three different conversion techniques are introduced based on the work of \citet{Brooks2003}, that transform the transdimensional ensemble into series of scalar, vector or tensor values – forms that are statistically analyzable. It will be shown that these transformed chains could be analyzed using standard convergence assessment techniques, without modification.  This paper is focusing on problems, where the objective is to reconstruct the heterogeneities of closed physical domains (geoscience imaging). The applicability of the introduced approaches is demonstrated on two different datasets, on a simple synthetic example, and on a field based tomographic fracture network modelling taken from \citet{Somogyvari2017}.

\section{An overview of the rjMCMC algorithm}
\label{sec:1}
The rjMCMC algorithm,introduced by \citet{Green1995}, is the transdimensional extension of the Metropolis-Hastings algorithm. The algorithm takes samples of the posterior probability distribution of the inverse problem, which is the multiplication of the prior and likelihood distribution according to Bayes' theorem.
Like other MCMC methods, rjMCMC is defined by an initial model and a transition kernel, the ruleset of generating the next sample in the chain. To converge and sample the posterior properly, the transition kernel should lead to an irreducible, aperiodic and positive recurrent (stationary, once it reached the target distribution) Markov chain \citep{gelman2013bayesian}.\\
The rjMCMC algorithm has an iterative structure with two phases in each iteration: an update phase and an evaluation phase. In the update phase the model is being perturbed by one of the following updates: parameter modification, birth of a new parameter or death of an existing one. The update is chosen randomly, with a predefined probability. Parameter modification is straightforward as it does not change the dimensionality of the problem. It is a simple Metropolis-Hastings update \citep{gelman2013bayesian}. The definition of the transdimensional updates could be more difficult, as a transdimensional mapping function has to be defined with the involvement of auxiliary variables (see \citet{Green1995}). In practice however, this can be avoided if the transdimensional updates are defined according to the nature of the investigated models. If the space of the model parameters is discretized, values of new parameters can be drawn from it for the addition step, and model features can be simply removed at deletion. In a fracture network model for example, new fractures could be added to the existing ones, at specified addition points \citep{Somogyvari2017}. In continuous parameter space, new parameters can be drawn from pre-defined probability distribution functions (see \citet{Sambridge2012} for example).\\
In the evaluation phase, the updated model realization is evaluated using the Metropolis-Hastings-Green (MHG) acceptance criterion \citep{Green1995}:
\begin{equation}
\alpha(\theta^*|\theta_n) =min\left[1,\frac{p(\theta^*)}{p(\theta_n)} \frac{L(\xi | \theta^*)}{L(\xi | \theta_n)} \frac{q(\theta_n|\theta^*)}{q(\theta^*|\theta_n)} |J|\right].
\end{equation}
The MHG acceptance criterion is composed from the  the ratio of the prior probabilities $\frac{p(\theta^*)}{p(\theta_n)}$, ratio of the likelihoods $\frac{L(\xi | \theta^*)}{L(\xi | \theta_n)}$, the ratio of the proposal probabilities $\frac{q(\theta_n|\theta^*)}{q(\theta^*|\theta_n)}$ and the Jacobian of the underlying implicit transformations. The ratios are calculated between the updated model $\theta^*$ and the model of the previous iteration $\theta_n$.
The value of the MHG ratio is the acceptance probability of the updated model. In practice it is compared with a random number $ \beta$ drawn from a uniform distribution  between 0 and 1. If $ \beta < \alpha$ the updated model is accepted and $ \theta_{n+1} = \theta^*$, otherwise the update is rejected and $ \theta_{n+1} = \theta_n$.\\
For the calculation of the proposal ratio, the probability of the reverse update is required, hence all updates have to be reversible. The role of the Jacobian is to preserve the stationarity of the Markov chain while transforming the probability distribution from one dimensionality to another when complex transition functions are used. However if the parameter spaces are fully discretized (and the used probability distribution functions are not continuous) the value of the Jacobian will always be one \citep{Denison2002}. Independent model proposals have the same property \citep{Agostinetti2010,Dosso2014}. In this paper problems with fully discretized parameter space are presented, but the introduced approaches are applicable for continuous space problems without any restriction.\\
The accepted realizations get stored in the ensemble, the final product of the MCMC inversion. Then the next iteration starts with the updated model. If an update is rejected, the chain stays at $\theta_n$. The resulted Markov chain provides an approximated sample of the posterior probability distribution of the inverse problem.

\section{Handling the transdimensional ensemble}
MCMC simulation generates thousands of model realizations to provide an estimate of the posterior of the inverse problem. The early part of the MCMC sequence is biased by the chosen initial model, and is not necessarily part of the solution of the inverse problem, i.e. the posterior distribution. In practice it is discarded as the burn-in and not considered part of the final result. During the burn-in, the models are converging toward the target stationary distribution of the Markov chain \citep{Brooks2011}. To find the point in the sequence where the convergence is completed is the task of convergence assessment. Typically it is achieved by monitoring the evolution of the statistics of the model parameters to have an indication if the MCMC chain has reached a stationary distribution. A well designed rjMCMC converges toward the stationary distribution if the chain is run for sufficient time. This time could be optimized by design choices of the Markov chain, for example by proposing a transition kernel that explores the parameter space faster.
\begin{figure}[h!]
\includegraphics[width=\textwidth]{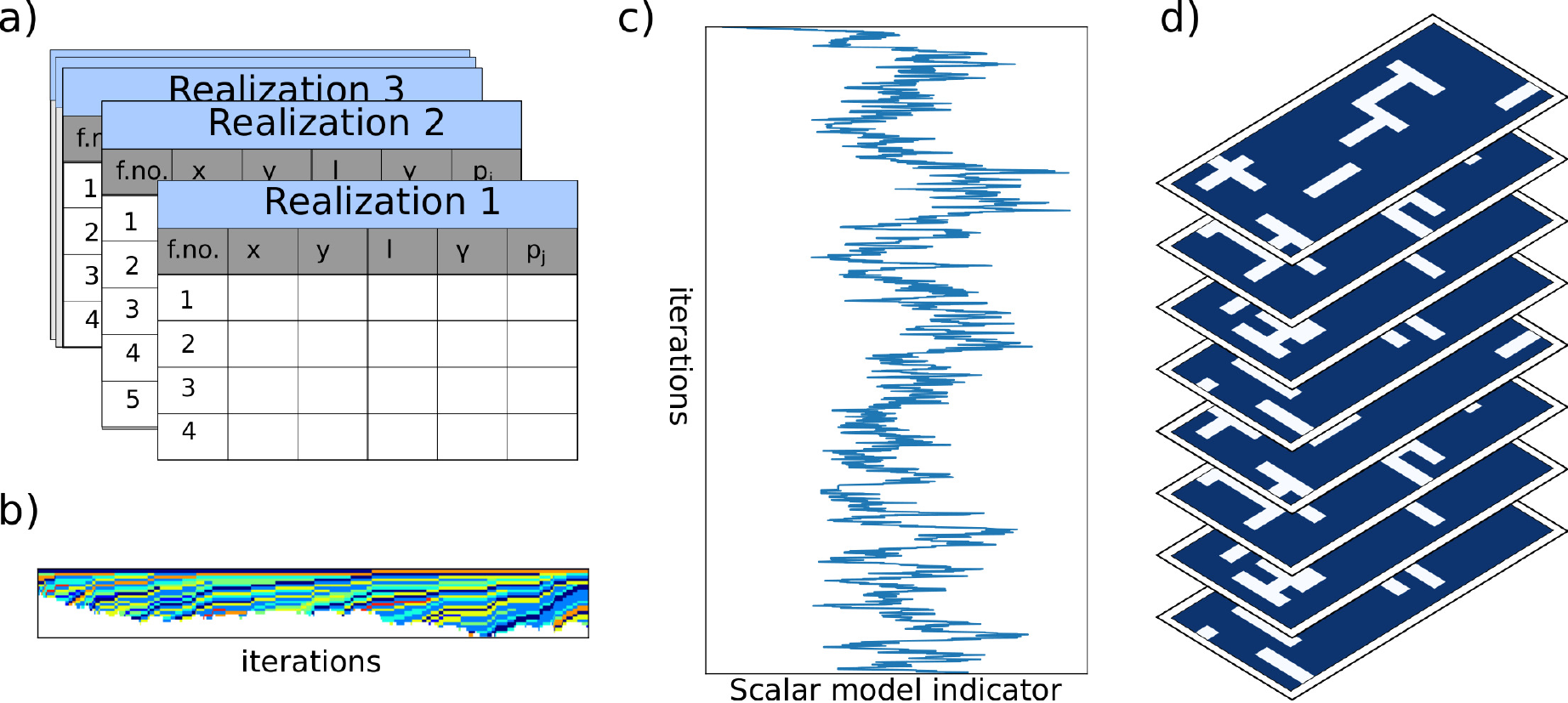}
\caption{Overview of the data conversion methods used to transform the transdimensional ensemble to a statistically analysable form. a) Data structure of a transdimensional geoscience imaging example with feature parameters: x,y coordinates, l,$\gamma$ size and orientation and $p_j$ physical properties. b) The ensemble converted into a series of substitute scalars. c) 2-D geological projection of the models in the transdimensional ensemble via rasterization.}
\label{conversion}
\end{figure}

\noindent In the case of transdimensional inversion, individual parameters cannot be followed through the sequence, as they can disappear and reappear at any point in the chain. This makes any statistical analysis and convergence assessment very challenging. In practice usually the evolution of the misfits is inspected to decide if the chain is converging. \citet{Somogyvari2017} proposed to monitor the changes in the mean of the Markov chain, and terminate the simulation when this mean does not change any more. Still, these solutions do not use any further statistical information about the transdimensional Markov chain.\\ 
As a first solution to assess the convergence of transdimensional chains, scalar hyperparameters of the models can be chosen and their evolution can be monitored \citep{Brooks2003}. This could be for example the number of components in a mixture, or the number of parameters within the model (see section 5.4).\\
In the following, we present three approaches that treat transdimensionality by transforming the MCMC sequence to forms that can be analysed by classic convergence assessment techniques (Fig. \ref{conversion}). Model realizations can be substituted by indicator parameters that uniquely identify the models. This concept is from \citet{Brooks2003}, where models with different dimensions were converted to scalar values, while preserving the statistical properties of the ensemble. 
Each model realization in the MCMC sequence can be described as the vector:
\begin{equation}
\theta_i=[\theta_i(1),...,\theta_i(k)],
\end{equation}
where $k$ notes the dimensionality of model $\theta_i$. In a transdimensional ensemble the length of these vectors could change from sample to sample.
\cite{Brooks2003} defines the following expression to convert $\theta_i$ to an equivalent scalar value, the scalar model indicator:
\begin{equation}
\nu(\theta_i)=1+\sum_{j=1}^k\theta_i(j)d^{j-1}\in\{1,...,d^k\}.
\label{SMI}
\end{equation}
This transformation is a bijection, where each model gets its own unique scalar model indicator. The detailed description and the inverse function of this bijection can be found in \citet{Brooks2003}. $d^k$ represents the cardinality of the problem, the number of possible models. With this transformation, the ensemble is reduced to one dimension, an easily analyzable form.
The geoscience imaging problems discussed in this paper are of variable selection problems, where a parameter is either present in a model or not. In this case, the cardinality of a model with $k$ dimension is $ d^k=2^k$. \\
In geoscience applications the variable selection problem is often more complicated, as it is not individual model parameters that are added or removed from the models, but model features with multiple parameters, such as geological layers, or fractures. A model fracture for example require parameters about its location, about its size and about its physical properties. Calculating the scalar model indicator in these cases could lead to values that are dominated by one of the parameter types. For example, coordinates of the model features could have significantly larger values than physical properties, such as fracture permeability. The same goes with parameters of different units. To overcome this problem, the calculation of a separate scalar model indicator for each parameter type within the model is recommended, to form a scalar model indicator vector. For each type of parameter, the scalar model indicator is calculated separately by eq. 3, then these scalars are put together in a vector size equal to the number of different parameter types. This modified scalar model indicator vector will have the same size for each model in the rjMCMC chain.\\
There is a third solution for geoscience imaging applications, where the model realizations represent a closed volume of the geological environment, a part of the subsurface. Hence, all realizations in the ensemble could be projected to a 2-D or 3-D geological profile. The projections are of same dimension if they created over the same grid, and can be statistically analyzed. The projection is typically performed via rasterization, by defining a fine grid over the area of interest then assigning the properties of the model pixel-by-pixel (or voxel-by-voxel). The grid cells could contain multiple parameters. This transformation is not necessarily a bijection, only if a very high resolution grid is used. The result of the conversion is a tensor, which is referred to as the geological projection. The main advantage of this solution is that the results are suitable for visualization directly. \citet{Bodin2009} used this transformation to visualize the mean of an ensemble of seismic velocity distributions, while \citet{Somogyvari2017} and \citet{Jimenez2016} used similar projections to generate probability maps of the investigated geological features.\\
Note that this approach is more straightforward when dealing with models of continuous features (like velocity distributions in seismic tomography), but we will show in the following that it is well applicable for discrete models as well. In the case of fracture modelling, this approach is similar to converting discrete fracture models to equivalent continuum models\citep{Illman2003,sahimi2011flow}. While the DFN models are better for simulations during the inversion, as DFN-based forward models are very fast and efficient (by reducing the physical modeling to one or two dimensions \citep{Somogyvari2017}), continuum models are better for analysis and ensemble visualization purposes. Hence, this methodology benefits from the advantages of both approaches.

\section{Investigated synthetic datasets}
\begin{figure}[h!]
\includegraphics[width=0.5\linewidth]{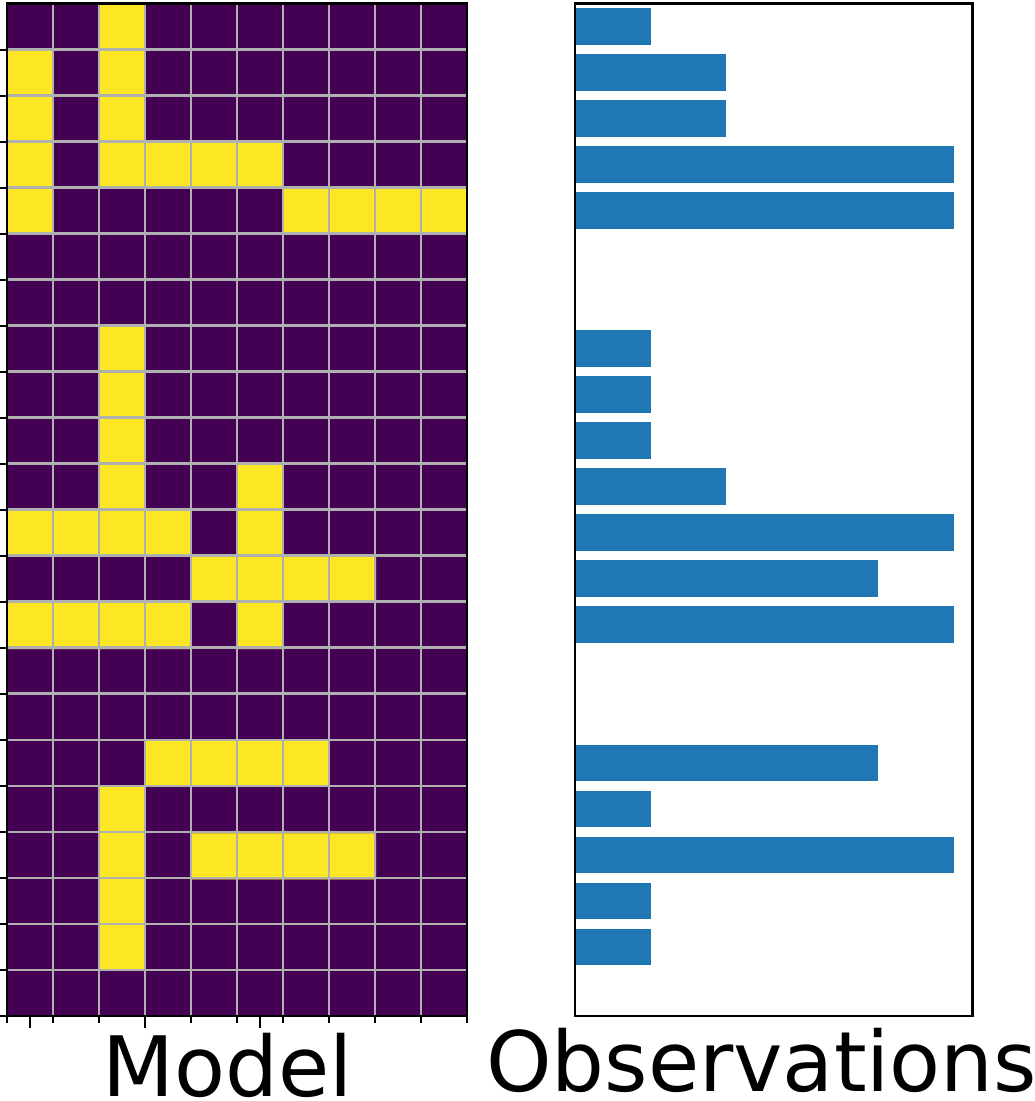}
\caption{Simplified DFN model and simulated observations in horizontal dimension. There is a similar set of observations in the vertical dimension too, which is not shown.}
\label{simple_model}
\end{figure}

In this paper, the proposed methodologies are tested on discrete fracture network models (DFN). DFNs are complex multidimensional models of fractured rocks that contain all relevant information of the fractured media in a simple data structure. The dataset contains the coordinates of the individual fractures, their length and orientation and the relevant physical properties such as fracture transmissivity, aperture or thermal and mechanical properties \citep{Somogyvari2017}.\\
Two separate datasets were used: a simplified, reduced DFN model for early tests, and a field-based realistic one. The simplified model has the following restrictions: fractures can only be vertical or horizontal, every fracture has the same length, and fractures can only positioned according to a predefined coarse grid. The model represents fractures as filled cells of this grid. No physical fracture properties are considered, only the geometry of the fracture network. Hence the simulated observations are simply taken as a measure on the geometry, as shown in Figure \ref{simple_model}.

\noindent For simulated observations, the number of filled grid cells is counted in each row and each column of the model (which is the table with the number and location of the fractures). This is equivalent to a simplified tomographic experiment with a dense source-receiver configuration and strictly straight signal trajectories, where the travel times are calculated as the sum of the time the signal requires to travel through the grid cells. The resulting two observation vectors together contain enough information for the reconstruction of the geometry of the system. These types of datasets are suitable for fast forward and inverse simulations and thus ideal subjects for method developments. For the analysis, multiple rjMCMC chains are run, using the simulated observations as input data. The generated sequences are 20000 iterations long, which are produced in a matter of minutes on a laptop.
\\
The methodologies are also tested on a field-based dataset of fracture networks. Data from \citet{Somogyvari2017} was used, where synthetic tracer tomography experiments were interpreted using rjMCMC. Tracer tomography is a novel tool of hydrogeological exploration, which is capable of reconstructing the distribution of the hydraulic properties of subsurface aquifers \citep{Brauchler2013,Jimenez2016,Zhu2009}. Tracer tomography uses subsequent injections of a tracer substance into the investigated aquifer. The travel times of the tracer between different source-receiver combinations provide information of the differences in transport properties in different aquifer parts.\\
The experiment was simulated on a synthetic fracture network which has been mapped from the Tschingelmad outcrop from the central Alps \citep{Ziegler2014}. Conservative tracer transport was modelled with a fast 1-D advection-diffusion forward model developed by \citet{Jalali2013}. The simulated experiments were interpreted using an rjMCMC sampler. A more detailed description of the used algorithm can be found in \citet{Somogyvari2017}. The inversions were run up to 100000 iterations on multiple chains using different random generated initial solutions. In contrast to the simple model, for these simulations up to 3-4 days were required on a high-performance office computer. No quantitative convergence assessment was done in \citet{Somogyvari2017}, rather chain lengths were set by visual inspection of the RMS error values. For the final ensemble the first $50\%$ of the chains were discarded as burn-in similarly to other practical studies \citep{Brooks2011}. Because the models within the simulated chains are strongly correlated, the sequences were thinned by keeping only every 100th member.

\section{Convergence assessment methods}
In this section, three different methods of convergence assessment are applied on the transdimensional ensemble of the simplified DFN model. The ensemble is converted to a scalar model indicator, a modified scalar model indicator and to a geologic projection and the methods are applied to each conversions. The goal is not to select the best convergence method, only to show that with the conversion of the transdimensional ensemble it is possible to assess convergence, and different conversions could lead to comparable results (here especially the modified scalar indicator and the geologic projection). Implementations of convergence assessment from the CODA package were used, coded in the pymc python package \citep{best1996coda,Fonnesbeck2013}. Note that all the presented equations are applicable on any type of model indicator. For simplicity they are called models in the following with notation $\theta$. Note that this $\theta$ differs from the ones used in the third section.

\subsection{Autocorrelation}
Although it is not a convergence assessment technique in a classical sense, calculating the autocorrelation of the Markov chain could provide valuable information about its behavior. The rjMCMC algorithm generates new model realizations via small updates to the previous one. This means that subsequent elements of the MCMC ensemble are strongly correlated. In a stationary Markov chain the autocorrelation does not change in time. Hence, autocorrelation can be used to verify stationarity. Also, the shape of the autocorrelation gives an indication on the efficiency of the MCMC. If the correlation decays quickly the sampler moves faster in the parameter space \citep{Link2012}. Strong autocorrelation could indicate the need for sequence thinning within the chain - to get closer to the ideal state of independent chain members.\\
The k-lag autocorrelation for a Markov chain can be calculated as:
\begin{equation}
\rho_k=\frac{\sum\limits_{i=1}^{n-k}\left(\theta_i-\bar{\theta}\right)\left(\theta_{i+k}-\bar{\theta}\right)}{\sum\limits_{i=1}^n(\theta_i-\bar{\theta})^2},
\end{equation}
where $\theta_i$ is the $i$-th model realization within the investigated Markov chain with $n$ length. $\bar{\theta}$ is the average of the models over the full chain.
\begin{figure}[h!]
\includegraphics[width=\textwidth]{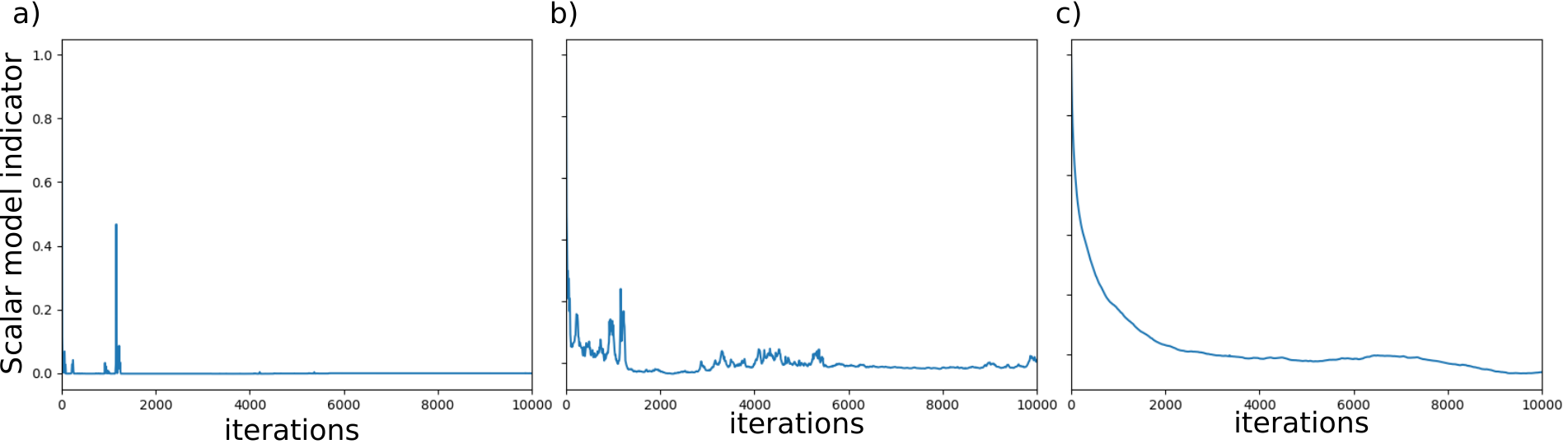}
\caption{Autocorrelation of the simple model case using a) scalar model indicator b) modified scalar model indicator c) 2-D projection.}
\label{simple_autocorr}
\end {figure}

\noindent The autocorrelation plots in Figure \ref{simple_autocorr} are ideal to compare the statistical behavior of the different conversion methods. As the output of the autocorrelation of the different conversion methods is a different sized vector, the arithmetic mean of these autocorrelation vectors for each realization are taken, to make them comparable. It is possible to use other norms for averaging, to take the median or an extrema, which may be advantageous in other applications. In the presented examples the 2-D projections are very sparsely populated grids, with possible very high local variations in some cells, hence the simple average is favored in this study.\\
The autocorrelation of the classic scalar model indicator shows spikes and rapid changes in values hence the analysis of this curve is not informative. This smoothed out in the modified scalar case due to the averaging, and the general shape of the curve shows a decay with the lag, which is expected from the autocorrelation of a Markov chain. The geological projection provides the smoothest curve, because of the largest vector size. The curves in Fig.\ref{simple_autocorr} b and c show a similar decaying trend, but the 2-D projection indicates slower mixing. The analysis of the scalar model indicator is not informative in this sense.

\subsection{Geweke diagnostics}
\begin{figure}[h!]
\includegraphics[width=\textwidth]{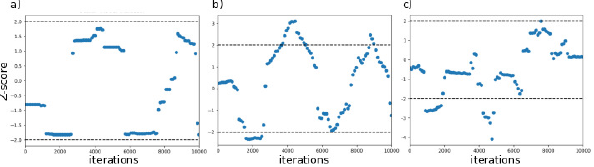}
\caption{Geweke diagnostics of the simple model case using a) classic scalar model indicator b) modified scalar model indicator c) 2-D projection.}
\label{simple_geweke}
\end {figure}

The idea behind the convergence diagnostic of \citet{Geweke1992} is simple: assuming that the Markov chain is long enough, the second half of the sequence should have already reached the stationary distribution. Hence it is possible to test the distribution of smaller intervals $A$ from the first part of the chain against the second half $B$ to find out if they are converged.
By calculating:
\begin{equation}
Z = \frac{\bar{\theta}_A-\bar{\theta}_B}{\sqrt{var(\theta_A)+var(\theta_B)}}.
\end{equation}
$Z$ should have approximately a standard normal distribution if the two subsamples are from the same distribution \citep{Cowles1999}. \citet{Geweke1992} recommends using spectral analysis to calculate the variances in the denominator.\\
During the shown analysis, the first half of the chain was split up to 100 smaller intervals, and they were tested against the whole second half of the chain. The results are shown in Figure \ref{simple_geweke}, with the indication of the 95 $\%$ probability level of significance of $|Z|$ at 1.96.

\noindent The Geweke plot of the scalar model indicator has its $Z$ scores within the limits all along the chain. Meanwhile the modified scalar model indicator and the geologic projection shows very similar attributes. The signs may differ, but the $Z$ score values are very close, showing indications of non-convergence at similar parts of the chain. This similarity is remarkable considering the differences in the two conversion techniques. These plots indicate that a significant part of the first chain half is not converged, showing major outliers until 5000 iterations (minor ones until 9000). The classic scalar model indicator does not provide this insight in this case.

\subsection{Gelman-Rubin diagnostics}
\citet{Gelman1992} developed convergence diagnostics for Gibbs sampling, which can be used for any type of MCMC method \citep{Cowles1999}. The main idea is that different Markov Chains should have similar statistical properties when they sample from the same target distribution. The method requires multiple MCMC chains with different starting points (ideally overdispersed along the posterior probability distribution of the problem). These chains are tested against each other. For the diagnostic the following statistical parameters of the chains are calculated: 
The within-chain variance is calculated as the average of the variance of each chain:
\begin{equation}
W=\frac{1}{C(T-1)}\sum\limits_{j=1}^C\sum\limits_{i=1}^T\left(\theta_{ij}-\bar{\theta}_j\right)^2,
\end{equation}
where $C$ is the number of chains and $T$ is their length. The between chain variance is calculated as the variance of the chain means:
\begin{equation}
B=\frac{T}{C-1}\sum\limits_{j=1}^C\left(\bar{\theta}_j-\bar{\bar{\theta}}\right),
\end{equation}
where $\bar{\bar{\theta}}$ is the mean over all models in all chains:
\begin{equation}
\bar{\bar{\theta}}=\frac{1}{C}\sum\limits_{j=1}^C\bar{\theta}_j.
\end{equation}
The variance of the target distribution is (over-) estimated from the samples with the weighted average of $W$ and $B$ (pooled posterior variance).
\begin{equation}
\widehat{Var}\left(\theta\right)=\left(1-\frac{1}{T}\right)W+\frac{1}{T}B.
\label{GR1}
\end{equation}
In the limit of $T\rightarrow\infty$ the variance of each chain approaches the target variance. It is possible to calculate the factor by which the scale of the current distribution of $\theta$ could be reduced by continuing the chain. This term is the scale reduction factor:
\begin{equation}
R=\sqrt{\frac{\widehat{Var}\left(\theta\right)}{\sigma^2}}.
\end{equation}
The value of $R$ approaches one as $T\rightarrow\infty$ and can be used as an indicator for convergence. The denominator of $R$ however is not known (the real variance of the target distribution, $\sigma^2$), so it must be estimated. To overestimate the value of $R$ (as smaller values indicate the convergence), $\sigma^2$ is underestimated by the within-chain variance, $W$ ($W$ is very likely an underestimation of $\sigma^2$ if the MCMC chain has not explored the complete posterior distribution.):
\begin{equation}
\hat{R}=\sqrt{\frac{\widehat{Var}\left(\theta\right)}{W}}.
\end{equation}
This expression is known as the potential scale reduction factor (PSRF). Note that this estimation could be further improved by accounting for the sampling variability (see \cite{Brooks1998a} for details).\\
If the value of $\hat{R}$ is large, that means further iterations could bring the chain closer to the target distribution. As the chain length increases, the potential scale reduction factor gets closer and closer to the value of one. \citet{Brooks1998a} recommended a cut-off value of 1.2 for $\hat{R}$ where the chains could be considered converged, but this study uses a more stringent 1.1 value which is more common in practice.
\begin{figure}[h!]
\includegraphics[width=\textwidth]{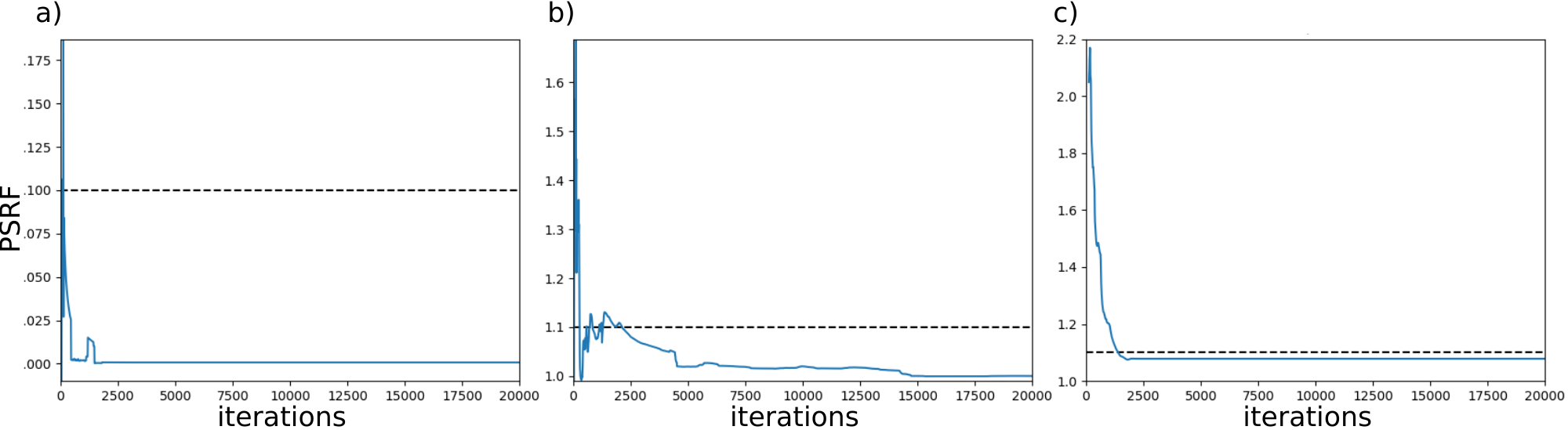}
\caption{Gelman-Rubin diagnostics of the simple model case using a) classic scalar model indicator b) modified scalar model indicator c) 2-D projection. The dashed lines represent the cut-off value of 1.1.}
\label{simple_GR}
\end {figure}

\noindent Figure \ref{simple_GR} shows the results of the Gelman-Rubin diagnostics of the simple model, with the 1.1 cut-off level indicated. Similarly to the previous two cases, the modified scalar model indicator and the geological projection give very similar curves, while the characteristic features of the classic scalar model indicator look damped and smaller in value. The other two figures show a rapid convergence within the first 2000 iterations. Both scalar representations show spikes on the plots, where the PSRF values suddenly drop, sometimes even to one. This is caused because the scalar conversions are more likely to provide random equalities between the chains at different points. While the other two approach the target value of 1, the PSRF of the 2-D projection does not. The reason for this is the existence of permanent differences between the chains, due to artifacts of MCMC modeling or by being trapped in a local minimum. This is very common in discrete fracture network inversion, as it is to be shown in the following section. Compared to the Geweke diagnostics, the point of convergence happens much earlier, indicating that different convergence assessment tests could give different results even on the same MCMC chain.

\subsection{Evolution of the dimensionality along the Markov chain}
\begin{figure}[h!]
\includegraphics[width=\textwidth]{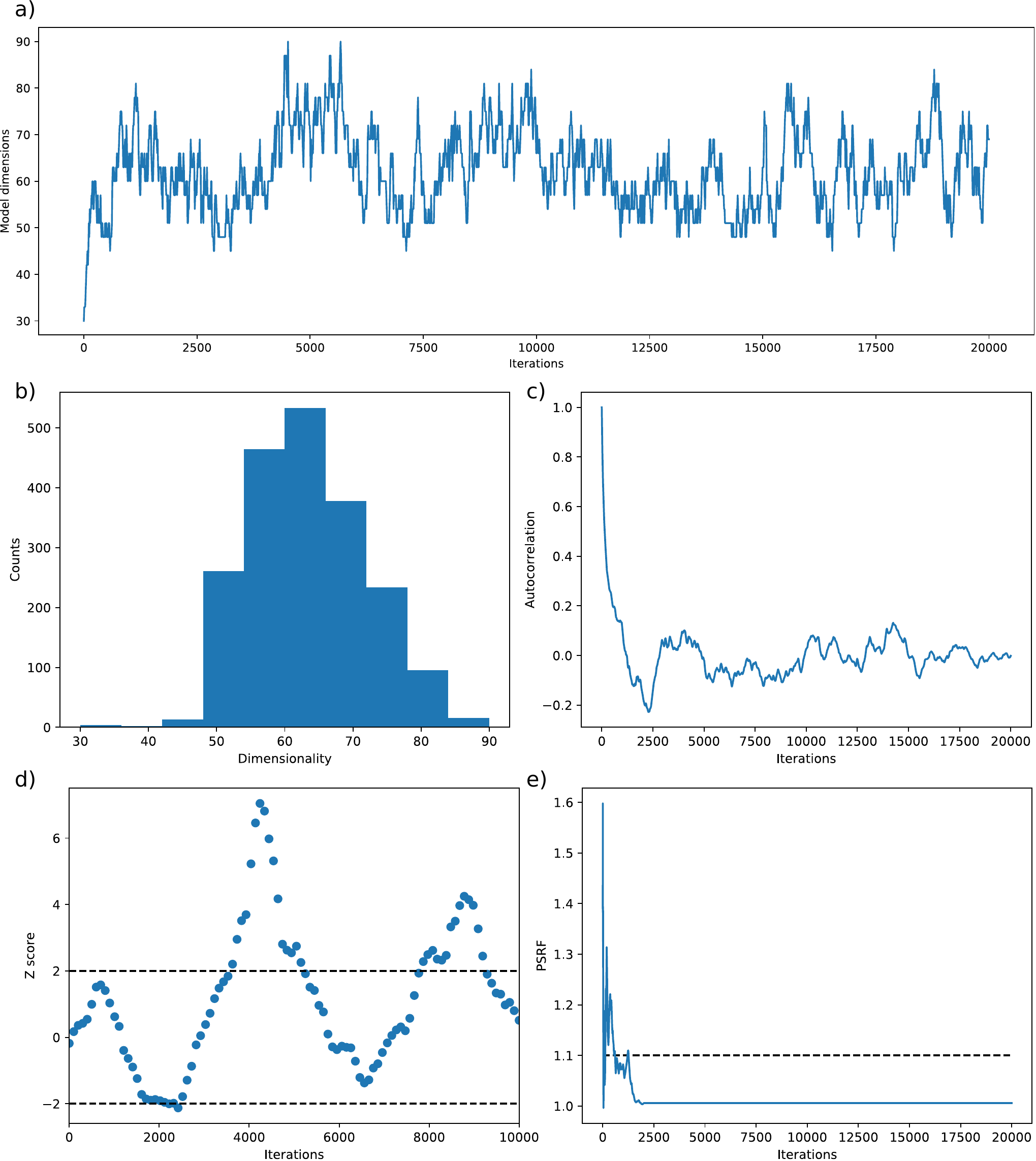}
\caption{Convergence analysis on dimensionality of the simple model case: a) evolution of the model dimensionality b) histogram of model dimensions c) autocorrelation d) Geweke diagnostics e) Gelman-Rubin diagnostics.}
\label{dim}
\end{figure}

In addition to the model conversion approaches presented here, it is possible to use an implicit quantity to extract information on the convergence of a transdimensional MCMC chain. When working with mixing problems with transdimensionality, it is a common way of convergence monitoring to follow the changes in number of components through the sequence \citep{Brooks2003}. Similarly to this, the changes in parameter numbers could also give an indication about the convergence. By following the variations in dimensionality it is easy to recognize that it has its own convergence behaviour, as it is shown on Fig. \ref{dim} a and b. Existing studies use this information to demonstrate the importance of the transdimensional approach \citep{Bodin2009,Jimenez2016}. Hence it is useful to analyze the convergence of the dimensionality with the presented convergence assessment approaches directly.

\noindent The three different diagnostic curves are very similar to the ones created with the conversion methods, showing that dimensionality has a significant impact on the convergence behaviour in this case. The convergence in dimensions occurs in parallel with the convergence in the parameter values.
\section{Analysis of the DFN ensemble}
\begin{figure}[h!]
\includegraphics[width=\textwidth]{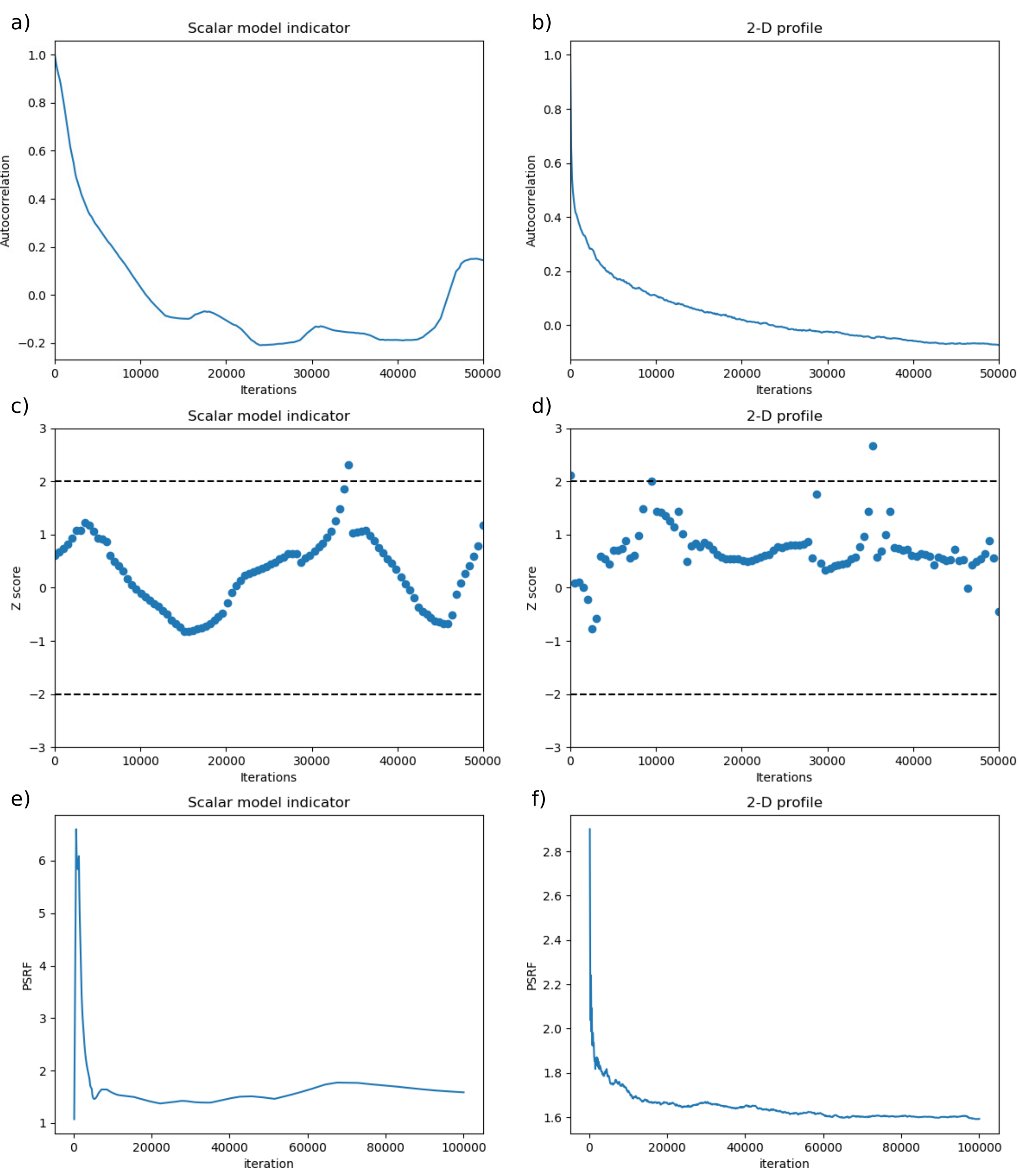}
\caption{Analysis of a field base transdimensional ensemble of reconstructed DFNs using tracer tomography data. a) Autocorrelation of the modified scalar model indicator of 5 independent chains, b) Autocorrelation of the 2-D projection of 5 independent chains, c) Geweke diagnostics of the modified scalar model indicator, d) Geweke diagnostic of the 2-D projection, e) Gelman-Rubin diagnostics of the modified scalar model indicator, f) Gelman-Rubin diagnostics of the 2-D projection.}
\label{DFN_analysis}
\end{figure}

In this section we apply the introduced methodology to a more complex, realistic dataset from DFN inversion from \citet{Somogyvari2017}. 
DFN models could consist of hundreds or thousands of discrete features, which could lead to very large scalar model indicator values. The calculation of eq. \ref{SMI} would use multipliers above $2^{100}-2^{1000}$ which could require special computational resources (as double precision 64 bit floating point numbers are limited at $1.7\cdot10^{308}\approx2^{1000}$). Using the modified scalar model indicator reduces the size of these numbers to a fraction, but still could be problematic with higher feature numbers. In our case only the classic scalar model indicator was non-computable using a standard office computer. The geological projection could be used in these cases.\\
One big difference compared to the previously presented simple model is the scale. The size of the rasterized DFN models are significantly larger. This means when calculating statistical properties pixel-by-pixel, the averaged final values will be much more smoothed. Also, large parts of the rasterized DFN models are zeros and does not have any significance when calculating the statistical properties, but they are damping the variances when the averages are calculated.\\
The results of the different conversion and diagnostic methods are presented on Figure \ref{DFN_analysis}. For the autocorrelation plots, we show 5 independent chains, for Geweke diagnostics one example chain, and for Gelman-Rubin diagnostics the PSRF calculated from the comparison of 5 chains. Note that the scalar model indicator conversion is not shown here, due to the clear advantage of the modified version.

\noindent The plots in Figure \ref{DFN_analysis} show similar diagnostics for both preferred conversion methods. At first look, the curves in the two autocorrelation plots strongly differ, mainly because of the averaging for the 2-D projection. Another important difference is the variability between different chains within the modified scalar vector indicator, which completely disappears for the 2-D projection. Still, the general trends of the two indicator types are similar, and averaging the 5 chains would lead to a very similar result. Note that these autocorrelation curves show longer correlation lengths than the simple model case. Here, models are generated via much smaller relative changes, making the subsequent chain elements strongly correlated - meaning that the exploration of the model parameter space is slow. This indicates the need for longer MCMC chains and stronger thinning for independent samples.\\
The two Geweke diagnostic plots show similarities, with outliers at the same parts of the chain. Indications of non-convergence are found in the sub 10000 and the 3-40000 range. These points however are easy to miss by using different segmentation during the analysis.\\
Gelman-Rubin diagnostics shows the largest difference in comparison to the simple model. Although the shapes of the curves are similar, the calculated values differ. Both curves saturate at values higher than one, which means that the between-chain variance ($B$ from eq. \ref{GR1}) does not converge to one, and the independent chains remain different. This could be explained by the existence of artifacts within the models that do not play a role in the inversion. These fractures could "freeze in" during a sequence, and be the source of permanent differences between separate chains. The inversion algorithm of \citet{Somogyvari2017} could also lead to different result DFN models in different model runs that are similar from the perspective of the used investigation method. Finding a better similarity measure between these models would probably resolve the offset in the diagnostic, but it would only work as a case specific solution. As a general solution, it is always possible to draw conclusions from the trends of the curves, and not to use pre-defined absolute cutoff values. Comparing with the geweke plots it could be said that the convergence of the rjMCMC chains finished after 30-40000 iterations. This is an earlier point than the one used in \citet{Somogyvari2017}, where the first half of the Markov chain was discarded. 

\section{Conclusions}
Converting the transdimensional ensemble of rjMCMC into a scalar model indicator or geological projection gives the opportunity to apply classic convergence assessment techniques directly. In this paper, it was demonstrated that different conversion approaches could preserve the original statistical properties of the ensemble, and could yield similar results regarding the convergence behavior.\\
The two types of scalar model indicators are usually the most convenient solutions to analyze ensembles with varying dimensions. They are easy to calculate, and reduce the convergence assessment problem to a handful of dimensions. In geoscience applications, where transdimensionality often originates from the number of features rather than parameters, using one scalar model indicator per model is not sufficient; one scalar value has to be calculated for each parameter type. This modified scalar model indicator yields similar diagnostic results as using a geological projection, using significantly fewer calculations. The geologic projection shows an advantage where the number of parameters is very large, and the scalar model indicator values becoming too large to handle. Because they can also be used for visualization purposes, they can also save an extra conversion step and be used directly for convergence assessment.\\
There is one question of convergence assessment not covered in this paper: if the sampling approximately captures the shape of the posterior. Making sure that an MCMC sampler visits all important part of the posterior probability function is a challenging task, and in theory would require to run the simulations infinitely long. With multi dimensional scaling the level of mixing could be demonstrated to a level \citep{Somogyvari2017,Caers2010}, however quantifying this property is still an open question even for classic MCMCs \citep{Robert1998,Brooks2011}. As a simple solution, the quick decay in the k-lag autocorrelation functions could be used as the indication for mixing, as the chain produces independent samples.\\
As unsupervised machine learning approaches are getting more and more popular today, transdimensional inversion could give a solution to create more flexible models, using only the observed data itself to constrain the model properties \citep{Sambridge2012}.
The use of rjMCMC however leads to one of the biggest questions of MCMC modeling: using one long chain or multiple short chains \citep{Brooks2011}. Multi-chain approaches can reduce computational times by parallel computing, thus getting more and more effective recently, but they are significantly more sensitive to the burn-in. The conversion methods could provide a tool to identify the point of convergence more accurately, and to minimize the bias from the initial model choice in these multi-chain applications. Convergence assessment is also important  when calculations are computationally more intensive, and only the production of shorter chains are possible.\\
The presented methodologies are very easy to implement, as all the utilized convergence assessment techniques are accessible from open software libraries and the conversions only require minimal modification of the investigated ensemble. 
More advanced test methods for convergence often require a special design of the Markov chain, thus are not as versatile as conversion. Debiased MCMC methods for example, use pairwise coupled MCMC chains that could provide more information on the convergence properties during simulation without any conversion \citep{Johnson1998,Johnson1996}.

\begin{acknowledgements}
This publication was financially supported by Geo.X, the Research Network for Geosciences in Berlin and Potsdam under the grant number SO_087_GeoX. This research has been also supported by the Deutsche Forschungsgemeinschaft (DFG) under grant 
CRC 1294 “Data Assimilation”. This is a pre-print of an article published in Mathematical Geosciences. The final authenticated version is available online at: https://doi.org/10.1007/s11004-019-09811-x
\end{acknowledgements}

\bibliographystyle{spbasic}      
\bibliography{library.bib}

\end{document}